\documentclass[aps,prl,superscriptaddress,reprint,showpacs,twocolumn,
preprintnumbers,amsmath,amssymb,floatfix]{revtex4-1}
\usepackage{graphicx}
\usepackage{epstopdf}
\usepackage{color}
\usepackage[latin1]{inputenc}
\usepackage{amssymb}
\usepackage{bm}
\usepackage{stmaryrd}
\usepackage{wasysym}
\usepackage{subfigure}

\begin{document}

\title{Superconductivity emerging from an electronic phase separation\\ in the charge ordered phase of RbFe$_2$As$_2$}
\author{E. Civardi}
\affiliation{Department of Physics, University of Pavia-CNISM,
I-27100 Pavia, Italy}
\author{M. Moroni}
\affiliation{Department of Physics, University of Pavia-CNISM,
I-27100 Pavia, Italy}
\author{M. Babij}
\affiliation{Institute of Low Temperature and Structure Research,
Polish Academy of Sciences, 50-422 Wroclaw, Poland}
\author{Z. Bukowski}
\affiliation{Institute of Low Temperature and Structure Research,
Polish Academy of Sciences, 50-422 Wroclaw, Poland}
\author{P. Carretta}
\affiliation{Department of Physics, University of Pavia-CNISM,
I-27100 Pavia, Italy}

\begin{abstract}
$^{75}$As, $^{87}$Rb and $^{85}$Rb nuclear quadrupole resonance
(NQR) and $^{87}$Rb nuclear magnetic resonance (NMR) measurements
in RbFe$_2$As$_2$ iron-based superconductor are presented. We
observe a marked broadening of $^{75}$As NQR spectrum below
$T_0\simeq 140$ K which is associated with the onset of a charge
order in the FeAs planes. Below $T_0$ we observe a power-law
decrease in $^{75}$As nuclear spin-lattice relaxation rate down to
$T^*\simeq 20$ K. Below $T^*$ the nuclei start to probe different
dynamics owing to the different local electronic configurations
induced by the charge order. A fraction of the nuclei probes spin
dynamics associated with electrons approaching a localization
while another fraction probes activated dynamics possibly
associated with a pseudogap. These different trends are discussed
in the light of an orbital selective behaviour expected for the
electronic correlations.
\end{abstract}


\maketitle

%\draft

%%%%%%%%%%%%%  TEXT  %%%%%%%%%%%%%%%%%%%%%%
%\section{\label{sec:intro}Introduction}

The parent compounds of high temperature superconducting cuprates
are emblematic examples of Mott-Hubbard insulators at half band
filling,\cite{MHub} where the large electron Coulomb repulsion $U$
overcomes the hopping integral $t$ and induces both charge
localization and an antiferromagnetic (AF) coupling among the
spins. Electronic correlations remain sizeable even when the
cuprates become superconducting and give rise to a rich phase
diagram at low hole doping levels characterized by the onset of a
charge density wave (CDW) which progressively fades away as the
doping increases \cite{Marc,Tranquada,Ghiringhelli,Hucker} and
eventually, in the overdoped regime, a Fermi liquid scenario is
restored. The comprehension of the role of electronic correlations
in iron-based superconductors (IBS)\cite{IBS} is more subtle. At
variance with the cuprates IBS are characterized by similar
nearest neighbour and next-nearest neighbour hopping integrals,
the parent compounds of the most studied families of IBS (e.g.
BaFe$_2$As$_2$ and LaFeAsO)\cite{Johnston} are not characterized
by half-filled bands and, moreover, in IBS the Fermi level
typically crosses five bands associated with the different Fe 3$d$
orbitals, leading to a rich phenomenology in the normal as well as
in the superconducting state.\cite{bands,Johnston} Moreover, even
if signs have been reported \cite{Mossba,Wang}, the evidence for a
charge order in the phase diagram of IBS still remains elusive.

Nominally, half band filling can be approached in BaFe$_2$As$_2$
IBS by replacing Ba with an alkali atom A=K, Rb or Cs, resulting
in 5.5 electrons per Fe atom.\cite{AFeAs1} Transport measurements
show that AFe$_2$As$_2$ compounds are metals\cite{AFeAs2} with
sizeable electronic correlations and it has been recently pointed
out that their behaviour shares many similarities with that of
heavy fermion compounds.\cite{Wu,AFeAs2} Indeed, the effective
mass progressively increases as one moves from BaFe$_2$As$_2$ to
AFe$_2$As$_2$,\cite{mass1} even if clear discrepancies in the
values derived by the different techniques are found depending on
their sensitivity to the electrons from a single band or from all
the five bands.\cite{mass2} de' Medici et al. \cite{Capone}
pointed out that if electronic correlations are sizeable, namely
$U/t$ is of the order of the unity, the local atomic physics
starts to be relevant and Hund coupling may promote the single
electron occupancy of Fe $d$ orbitals (i.e. half band-filling) and
decouple the interband charge correlations. Accordingly the Mott
transition becomes orbital selective\cite{Capone,Capone2} so that
while the electrons of a given band localize the electrons of
other bands remain delocalized, leading to a metallic behaviour
and eventually to superconductivity. This orbital selective
behaviour should give rise to markedly $k$-dependent response
functions \cite{Gull} and to a sort of $k$-space phase separation
of metallic and insulating-like domains. The point is, what
happens in the real space? Will one probe the sum of the
insulating and metallic response functions or should one detect  a
real space phase separation\cite{Emery} also in AFe$_2$As$_2$ IBS
\cite{Dagotto}, with different local susceptibilities ? More
interestingly, if electronic correlations become significant in
AFe$_2$As$_2$ one could envisage the onset of a charge
order\cite{DiCastro} as in the
cuprates.\cite{Marc,Tranquada,Ghiringhelli,Hucker}

Nuclear quadrupole resonance (NQR) and nuclear magnetic resonance
(NMR) are quite powerful tools which allow to probe the local
response function and charge distribution. Moreover, in NQR
experiments \cite{Abragam} the magnetic field, which often acts as
a relevant perturbation, is zero. Here we show, by combining
$^{75}$As and $^{87,85}$Rb NQR and $^{87}$Rb NMR measurements,
that in RbFe$_2$As$_2$ a charge order develops in the normal state
below $T_0\simeq 140$ K, possibly leading to a differentiation in
real space of Fe atoms with different orbital configurations.
Below $T_0$, $^{75}$As and $^{87}$Rb nuclear spin-lattice
relaxation rates ($1/T_1$)  show a power law behaviour, as it is
expected for a strongly correlated electron system and in good
agreement with $^{75}$As NMR results reported by Wu et
al.\cite{Wu}. However, at $T^*\simeq 20$ K we observe that a
fraction of $^{75}$As (or $^{87}$Rb) nuclei probes spin dynamics
characteristic of a system approaching localization while others
probe dynamics possibly associated with a metallic phase with a
pseudogap.\cite{pseudogap,Ding,Batlogg} Upon further decreasing
the temperature the volume fraction of the heavy electron phase
vanishes while the one of the metallic phase, which eventually
becomes superconducting below $T_c\simeq 2.7$ K, grows. Thus, we
present a neat evidence for a charge order in RbFe$_2$As$_2$ akin
to underdoped cuprates. The charge order favours a phase
separation into metallic and nearly insulating regions, which
could result from the theoretically predicted orbital selective
behaviour.\cite{Capone}

\begin{figure}[h!]
\vspace{5.5cm} \includegraphics{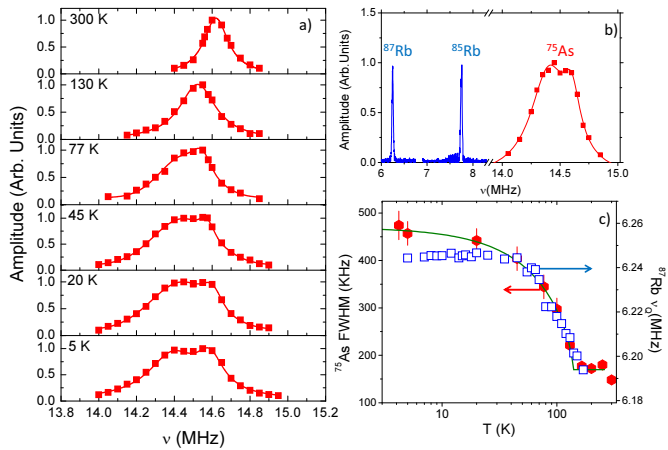}  \caption{(Color online) a) The
$^{75}$As NQR spectrum in RbFe$_2$As$_2$ is reported at different
temperatures between 5 K and 300 K. The red lines are best fits
with one or two (for $T< 130$ K) lorentzians. b) The merge of the
NQR spectra, associated with the $m_I=\pm 3/2\rightarrow \pm 1/2$
transition for $^{75}$As and $^{87}$Rb and with the $m_I=\pm
5/2\rightarrow \pm 3/2$ transition for $^{85}$Rb ($I=5/2$), is
shown for $T= 4.2$ K. The intensity of the three spectra has been
rescaled so that all three spectra have similar intensities. c)
The temperature dependence of the full width at half intensity
(FWHM) of $^{75}$As NQR spectra (red octagons, left scale, for the
plot with linear $T$ scale see the Supplemental Material
\cite{Supplem}) is shown together with the temperature dependence
of $^{87}$Rb $\nu_Q$ (blue squares, right scale). The green solid
line tracking the order parameter is a phenomenological fit of the
FWHM ($\Delta\nu_Q$) with $\Delta\nu_Q= 300(1-(T/T_0))^\beta +
170$ KHz, with $T_0=140$ K and $\beta\simeq 0.7$.} \label{Figspec}
\end{figure}

NQR and NMR measurements were performed on a RbFe$_2$As$_2$
polycrystalline sample with a mass of about 400 mg, sealed in a
quartz tube under a 0.2 bar Ar atmosphere in order to prevent
deterioration. The superconducting transition temperature derived
from ac susceptibility measurements turned out $T_c\simeq 2.7$ K,
in good agreement with previous findings \cite{Bukowski,muSR}.
Further details on the sample preparation and characterization are
given in the supplemental material.\cite{Supplem}

First of all we shall discuss the appearance of a charge order in
the FeAs planes of RbFe$_2$As$_2$, as detected by $^{75}$As NQR
spectra. For a nuclear spin $I=3/2$, as it is the case of
$^{75}$As and $^{87}$Rb, the NQR spectrum is characterized by a
single line at a frequency\cite{Abragam}
\begin{equation} \label{nqrspe}
\nu_Q= \frac{eQV_{ZZ}}{2h} \biggl(1+ \frac{\eta^2}{3}\biggr)^{1/2}
\,\,\, ,
\end{equation}
with $Q$ the nuclear quadrupole moment, $V_{ZZ}$ the main
component of the electric field gradient (EFG) tensor and $\eta$
its asymmetry $\eta= (V_{XX}-V_{YY})/V_{ZZ}$. Hence the NQR
spectrum probes the EFG at the nuclei generated by the surrounding
charge distribution. Above 140 K, $^{75}$As NQR spectrum
(Fig.\ref{Figspec}) is centered around 14.6 MHz, with a linewidth
of about 170 KHz, while $^{87}$Rb NQR spectrum is centered around
6.2 MHz with a width of about 20 KHz. The relatively narrow NQR
spectra confirms the good quality of our sample. We performed
density functional theory (DFT) calculations using Elk code in the
generalized gradient approximation\cite{Supplem} in order to
derive \textit{ab initio} the electric field gradient and NQR
frequency. For $^{75}$As and $^{87}$Rb we obtained
$(^{75}\nu_Q)_{\mathrm{DFT}}=14.12$ MHz and
$(^{87}\nu_Q)_{\mathrm{DFT}}=6.7$ MHz, respectively, in reasonable
agreement with the experimental values in spite of the significant
electronic correlations.\cite{Roser} This shows that DFT is still
able to provide a fair description of the system as far as it
remains a normal metal.

Upon cooling the sample below $T_0\simeq 140$ K significant
changes are detected in $^{75}$As NQR spectra (Fig.
\ref{Figspec}). The spectrum is observed to progressively broaden
with decreasing temperature and below 50 K one clearly observes
that the spectrum is actually formed by two humps nearly
symmetrically shifted with respect to the center
(Fig.\ref{Figspec}a).  The presence of two peaks in the $^{75}$As
NQR spectra has already been detected in different families of IBS
and associated with a nanoscopic phase separation in regions
characterized by different electron doping levels.\cite{Lang}
However, at variance with what we observe here, the two peaks
observed in other IBS do not show the same intensity \cite{Lang}
and the spectra show little temperature dependence, namely the
nanoscopic phase separation is likely pinned. Under both high
magnetic field and high pressure an asymmetric splitting of
$^{75}$As NMR spectrum was detected also in KFe$_2$As$_2$ which,
however, is absent in zero field (NQR).\cite{Wang} Here we observe
the emergence of an NQR spectrum which recalls the one expected
for an incommensurate CDW,\cite{Claude1,Slicht1,Claude2} which
causes a periodic modulation of the EFG at the nuclei and gives
rise to two symmetrically shifted peaks in the spectrum. The EFG
modulation could involve also the onset of an orbital
order\cite{orbitalord} or a structural distortion, possibly
coupled to the charge order. Although it is not straightforward
from our data to discriminate among these scenarios, it is clear
that we detect a symmetry breaking below $T_0$ to a low
temperature phase characterized by a spatial modulation of the
EFG, namely by a charge order.

\begin{figure*}[t]
\vspace{4.7cm}
\includegraphics{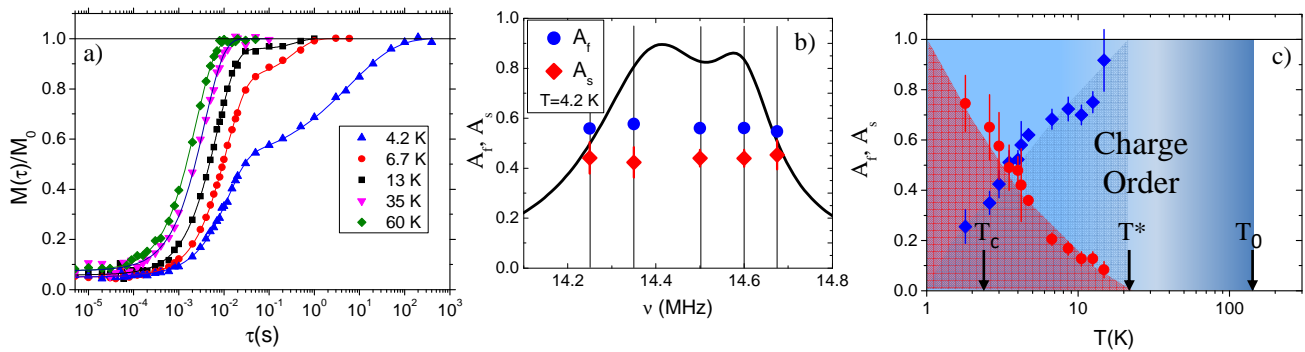} \caption{(Color online) a) The recovery of $^{75}$As
nuclear magnetization $M(\tau)$ (measured in NQR) is reported as a
function of the delay $\tau$ between a saturation radiofrequency
pulse sequence and the echo readout sequence for different
temperatures. The solid lines are the best fits according to Eq.2
in the text. b) The frequency dependence of the fraction of fast
$A_f$ and slow relaxing $A_s$ nuclei is reported as a function of
the irradiation frequency across $^{75}$As NQR spectrum. The black
solid lines is the best fit of the spectrum at $T=4.2$ K. c) The
temperature dependence of $A_f$ (blue) and $A_s$ (red) recorded on
the low-frequency peak of the $^{75}$As NQR spectrum.}
\label{Figrec}
\end{figure*}

$^{87}$Rb NQR spectrum does not show a significant broadening upon
decreasing the temperature but is characterized by a $\nu_Q$
which, at $T> T^*\simeq 20- 25$ K, shows a temperature dependence
similar to that of the $^{75}$As NQR spectra full width at half
maximum (FWHM), proportional to the charge order parameter
(Fig.\ref{Figspec}c). Below $T^*$ $^{87}$Rb $\nu_Q$ flattens and
deviates from $^{75}$As NQR FWHM. The fact that the NQR spectrum
of the out of plane $^{87}$Rb nuclei is less sensitive than
$^{75}$As one to the charge order is an indication that the order
develops in the FeAs planes and that the modulation of the EFG at
$^{75}$As nuclei should occur over a few lattice steps, otherwise
one should expect a splitting also of the narrow $^{87}$Rb NQR
spectrum. It is interesting to notice that at a temperature of the
order of $T^*$ an abrupt change in the uniaxial thermal expansion
occurs,\cite{Meingast} evidencing also a change in the lattice
properties.

Now we discuss the temperature dependence of the low-energy
dynamics probed by $^{75}$As and $^{87}$Rb $1/T_1$. The nuclear
spin-lattice relaxation rate was determined from the recovery of
the nuclear magnetization after exciting the nuclear spins with a
saturation recovery pulse sequence. The recovery of $^{75}$As
magnetization in NQR is shown in Fig.\ref{Figrec}a. One notices
that a single exponential recovery describes very well the
recovery of the nuclear magnetization at $T\geq 20$ K, as it can
be expected for a homogeneous system where all nuclei probe the
same dynamics. However, below $T^*\simeq 20$ K one observes the
appearance of a second component characterized by much longer
relaxation times. Namely, a part of the nuclei probes dynamics
causing a fast relaxation ($1/T_1^f$) and a part of the nuclei a
slow relaxation ($1/T_1^s$). Accordingly, the recovery was fit to
\begin{equation} \label{recovery}
M(\tau)=M_0\biggl[ 1 - f\biggl( A_f e^{-3\tau/T_1^f}+
A_se^{(-3\tau/T_1^s)^\beta}\biggr)\biggr] \,\,\, ,
\end{equation}
with $M_0$ the nuclear magnetization at thermal equilibrium, $A_f$
and $A_s$ the fraction of fast relaxing and slow relaxing nuclei,
respectively, $f$ a factor accounting for a non perfect saturation
by the radiofrequency pulses and $0.8\geq\beta\geq 0.3$ a
stretching exponent characterizing the slowly relaxing component.
As the temperature is lowered one observes a progressive increase
of $A_s$ with respect to $A_f$ and at the lowest temperature ($T=
1.7$ K), about 80\% of the nuclei are characterized by the slow
relaxation (Fig.\ref{Figrec}c). It is important to notice that in
RbFe$_2$As$_2$ Wu et al.\cite{Wu} (in NMR, not in NQR) did not
observe a clear separation of the recovery in two components as we
do here but they did observe deviations from a single exponential
recovery below 20 K which, however, were fitted with a stretched
exponential, likely yielding an average $1/T_1$ value between
$1/T_1^s$ and $1/T_1^f$. Remarkably also $^{87}$Rb NMR $1/T_1$
clearly shows two components below 25 K and just one
above.\cite{Supplem}

\begin{figure}[h!]
\vspace{9.5cm} \includegraphics{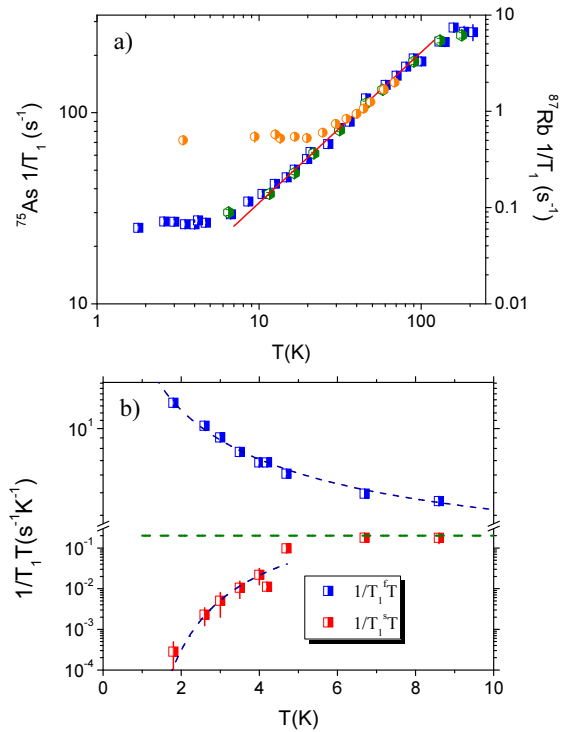} \caption{(Color online) a) The temperature
dependence of $^{75}$As NQR $1/T_1$ in RbFe$_2$As$_2$, for $T\geq
20$ K, and of the fast relaxation rate $1/T_1^f$, for $T<20$ K,
are reported for an irradiation frequency centered at the
low-frequency peak (blue squares) and for an irradiation frequency
centered at the high frequency peak (green circles). The red solid
line is a best fit to the data between 20 and 100 K with a power
law characterized by an exponent $b=0.79$. $^{87}$Rb NMR $1/T_1$
(orange circles) in RbFe$_2$As$_2$ is reported between 3.5 and 70
K, for an external magnetic field $H=7$ Tesla. b) The temperature
dependence of $^{75}$As NQR $1/T_1^f T$ (blue squares) and
$1/T_1^s T$ (red circles) in RbFe$_2$As$_2$ is reported. The
dashed line at the bottom is the best fit according to an
activated behaviour with an energy gap $E_g=17\pm 0.9$ K. The
dashed line at the top is the behaviour expected according to
Moriya SCR theory (see text). The dashed horizontal line shows
schematically the Korringa-like behaviour expected for an
uncorrelated metal. } \label{FigT1}
\end{figure}

$^{75}$As $1/T_1$ was measured both on the high frequency and on
the low-frequency shoulder of the NQR spectrum and it was found to
be the same (Fig.\ref{FigT1}a) over a broad temperature range.
Moreover, at $T=4.2$ K we carefully checked the frequency
dependence of $T_1^f$, $T_1^s$, $A_f$ and $A_s$ and found that
neither the two relaxation rates nor their amplitude  vary across
the spectrum (Fig.\ref{Figrec}b, see also
Ref.\onlinecite{Supplem}). This means that nuclei resonating at
different frequencies probe the same dynamics which implies that
the charge modulation induced by the charge order has a nanoscopic
periodicity.\cite{Lang} One could argue that the two components
are actually present at all temperatures but that they arise only
at low temperature once nuclear spin diffusion\cite{spindiff} is
no longer able to establish a common spin temperature (i.e. a
common $T_1$) among the nuclei resonating at different
frequencies. However, we remark that since the nuclear spin-spin
relaxation rate ($1/T_2$) is constant \cite{Supplem} and the width
of the NQR spectrum is nearly constant below 40 K
(Fig.\ref{Figspec}c) the poor efficiency of nuclear spin diffusion
should not vary, at least for $T\leq 40$ K. Hence, the appearance
of different relaxation rates below $T^*$ should arise from a
phase separation causing a slight change in the average electronic
charge distribution causing little effect on the NQR  spectra (see
Fig. \ref{Figspec}) but a marked differentiation in the low-energy
excitations \cite{Gull}, which starts to be significant at low
temperature once the effect of electronic correlations is
relevant.

One has to clarify if the relaxation mechanism is magnetic, driven
by electron spin fluctuations, or quadrupolar, driven by EFG
fluctuations, typically induced by CDW amplitude and phase
modes.\cite{Claude2} In order to clarify this point we measured
the ratio between $^{87}$Rb and $^{85}$Rb $1/T_1$ (fast component)
at a few selected temperatures below 25 K. The ratio
$^{87}(1/T_1)/^{85}(1/T_1)= 12 \pm 1$, in good agreement with the
ratio between the square of the gyromagnetic ratios of the two
nuclei $(^{87}\gamma/^{85}\gamma)^2= 11.485$, showing that the
relaxation is driven by the correlated spin fluctuations and not
by charge fluctuations associated with CDW excitations. Since
$^{75}$As shows a temperature dependence of the relaxation
analogous to the one of $^{87}$Rb (Fig.\ref{FigT1}a) we argue that
also $^{75}$As $1/T_1$ is driven by spin fluctuations. Thus we can
write that
\begin{equation}
\frac{1}{T_1}=\frac{\gamma_n^2}{2\hbar} k_B T\frac{1}{N}
\sum_{\vec{q}}
|A_{\vec{q}}|^2\frac{\chi''(\vec{q},\omega_0)}{\omega_0} \, ,
\end{equation}
with $|A_{\vec{q}}|^2$ the form factor giving the hyperfine
coupling with the collective spin excitations at wave-vector $\vec
q$, and $\chi''(\vec{q},\omega_0)$ the imaginary part of the
dynamic susceptibility at the resonance frequency $\omega_0$.

Now we turn to the temperature dependence of $1/T_1$ above
$T^*\simeq 20$ K and of $1/T_1^s$ and $1/T_1^f$ below that
temperature. Above $T^*$ $1/T_1$ increases with a power law
$1/T_1= aT^b$, with $b=0.79\pm 0.01$ for $^{75}$As,  and flattens
around $T_0\simeq 140$ K (Fig.\ref{FigT1}a), in very good
agreement with the results reported by Wu et al.\cite{Wu} from
$^{75}$As NMR. Notice that $T_0$ corresponds to the temperature
below which we start to observe a significant broadening of
$^{75}$As NQR spectrum. Hence, the power law behaviour of $1/T_1$
seems to arise from the onset of the charge order.
%A power law
%exponent close to 0.7 was observed in heavy fermion compounds with
%AF spin correlations close to a quantum critical point (QCP), as
%CeCu$_{5.9}$Au$_{0.1}$.\cite{CeCu}

%

Below $T^*\simeq 20$ K $1/T_1^f$ deviates from the power law
behaviour and progressively flattens on decreasing temperature
(Fig.\ref {FigT1}a). The same behaviour is detected for $^{87}$Rb
NMR $1/T_1$, although the flattening starts at a higher
temperature, suggesting that $T^*$ might be field dependent. On
the other hand, $1/T_1^s$ gets progressively longer as the
temperature is lowered and follows an activated trend with an
energy barrier $E_g=17 \pm 0.9$ K.

The behaviour of $1/T_1^f$ is characteristic of a system
approaching a QCP where localization occurs. In fact, from Moriya
self-consistent renormalization (SCR) approach for a quasi-2D
system with AF correlations, one should have $1/T_1=
T\chi(Q)$,\cite{Moriya1,Franziska} with $\chi(Q)$ the static
susceptibility at the AF wave-vector. In the proximity of the QCP
$\chi(Q)\sim ln(1/T)/T$, leading to a weak logarithmic divergence
of $1/T_1\sim ln(1/T)$ for $T\rightarrow 0$, while at higher
temperature $\chi(Q)$ should show a Curie-Weiss behaviour,
yielding a nearly flat $1/T_1$, as we do observe in RbFe$_2$As$_2$
(Fig.\ref{FigT1}a). The corresponding behaviour of $1/T_1^fT$ is
reported in Fig.\ref{FigT1}b.

On the other hand, $1/T_1^sT$, corresponding to the relaxation
rate of the majority phase at low temperature, shows the opposite
trend (Fig.\ref{FigT1}b), decreasing upon cooling. Being the
system metallic at low temperature, the deviation of $1/T_1^sT$
from the constant Korringa-like behaviour \cite{Abragam} expected
for a metal should possibly be associated with the opening of a
pseudogap, similarly to what one observes in the underdoped regime
of the cuprates,\cite{pseudogap,Ding,Batlogg} and in agreement
with theoretical predictions for hole-doped IBS.\cite{Gull}

In conclusion, our results show that, akin to cuprates, a charge
order develops also in the normal state of IBS when the electronic
correlations are sizeable. Accordingly, the presence of a charge
order appears to be a common feature in the phase diagram of
cuprate and iron-based superconductors and could play a key role
in determining the superconducting state
properties.\cite{DiCastro,Caprara} Moreover, we observe a local
electronic separation in two phases characterized by different
excitations which could possibly be explained in terms of the
orbital selective behaviour\cite{Capone} predicted for IBS.
Finally we remark that the occurrence of an electronic phase
separation is theoretically supported by a recent study of the
electron fluid compressibility. \cite{Compress}

\section*{acknowledgments}
Massimo Capone is thanked for useful discussions. The Sezione INFN
di Pavia is acknowledged for granting the computing time necessary
to perform DFT calculations. This work was supported by
MIUR-PRIN2012 Project No. 2012X3YFZ2.

\section*{Supplementary Material }

\section{I. Sample Synthesis and Characterization}

A polycrystalline sample of RbFe$_2$As$_2$ was synthesized in two
steps.\cite{Bukowski} First, RbAs and Fe$_2$As precursors were
prepared from stoichiometric amounts of rubidium, arsenic and
iron. The components were mixed and heated in evacuated and sealed
silica tubes at 350 C and at 800 C, respectively. Then, the
obtained RbAs and Fe$_2$As were mixed together in a molar ratio
1:1, pressed into pellets and placed in an alumina crucible and
sealed in an evacuated silica ampoule. The sample was annealed at
650 C for three days, ground and annealed for another three days
at the same temperature. It should be emphasized that the
annealing temperature and time are crucial parameters. The
annealing at higher temperature or extended annealing time causes
decomposition of the compound.

%%%%%%%%%%%%%%%%%%%%%%%%%%%%%%
\begin{figure}[h!]
\vspace{5.6cm} \includegraphics{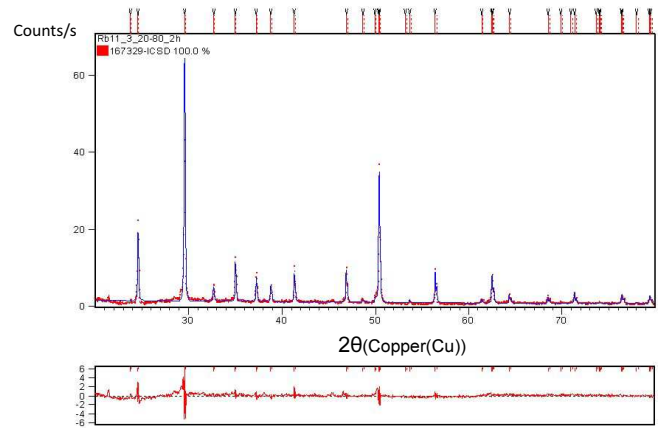} \caption{RbFe$_2$As$_2$2 room
temperature X-ray diffraction pattern.} \label{XRD}
\end{figure}
%%%%%%%%%%%%%%%%%%%%%%%%%%%%%%%%%

\begin{figure}[h!]
\vspace{6.5cm} \includegraphics{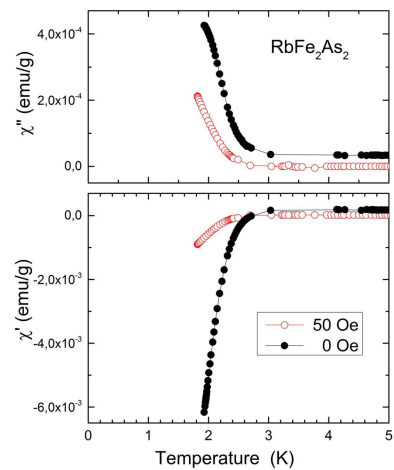} \caption{Temperature dependence of the
real ($\chi'$) and imaginary part ($\chi"$) of the ac
susceptibility close to $T_c$ in RbFe$_2$As$_2$ in zero field and
for a static external magnetic field of 50 Oe.} \label{chiac}
\end{figure}

The phase purity was checked by X-ray powder diffraction. The
diffraction lines (shown in Fig.\ref{XRD}) can be indexed with a
tetragonal ThCr$_2$Si$_2$ type unit cell with lattice parameters
$a=3.871$ \AA\, and $c= 14.464$ \AA\,, in good agreement with
those reported in Ref.\onlinecite{Bukowski}. The ac-susceptibility
measurements were performed on heating with an ac field of 10 Oe
at 1111 Hz. The real component (Fig.\ref{chiac}) reveals the onset
of diamagnetism and of bulk superconductivity below $T_c\simeq
2.7$ K.\cite{Bukowski}

%%%%%%%%%%%%%%%%%%%%%%%%%%%%%%
%%%%%%%%%%%%%%%%%%%%%%%%%%%%%%%%%

\section{II. DFT calculations}

First-principles DFT calculations of the electronic structure were
performed using the full-potential linearized augmented plane-wave
method as implemented in the Elk package~\cite{elk}. For the
exchange-correlation functional we used the generalized gradient
approximation of Perdew, Burke, and Ernzerhof ~\cite{gga_pbe}. The
atomic positions used in the calculation are those obtained from
room temperature x-ray diffraction. In order to calculate the
electric field gradient (EFG) tensor components $V_{ij}^\alpha$ we
solved the Poisson equation for the charge distribution  to
determine the electrostatic potential $\varphi$ and derived
$V_{ij}^\alpha$ from
\begin{equation}\label{efg}
V_{ij}^\alpha=\left.\dfrac{\partial^2 \varphi}{\partial
\mathbf{r}_i\partial\mathbf{r}_j}\right|_{\mathbf{r}_\alpha}
\mbox{ ,}
\end{equation}
where $\alpha$ runs over the nuclei at $\mathbf{r}_\alpha$. Since
the EFG tensor is extremely sensitive to the charge distribution a
well converged basis set is needed to grant the convergence with
respect to the EFG tensor components. We used muffin tin radii of
2.6 $a_0$ for Rb and 2.4 $a_0$ for Fe and As, with
$R^{MT}_{min}\times max(|k|)=9$, where $R^{MT}_{min}$ is the
smallest muffin tin (MT) radius inside the MT spheres and $|k|$
the magnitude of the reciprocal space vectors. We choose 9 for the
cut off of the angular momentum quantum number in the lattice
harmonics expansion inside the MTs. Reciprocal space was sampled
with the Monkhorst-Pack~\cite{mhgrid} scheme on a
$24\times24\times24$ grid. A smearing of 270 meV was used within
the Methfessel-Paxton scheme~\cite{paxton} and convergence of the
EFG components with respect to all these parameters has been
carefully checked. Once the EFG tensor components are known the
NQR frequency at $^{75}$As and $^{87}$Rb can be calculated from
Eq. 1 in the main article. The obtained values,
($^{75}\nu_Q$)$_{\textrm{DFT}}=$ 14.12 MHz and
($^{87}\nu_Q$)$_{\textrm{DFT}}=$ 6.7 MHz are in good agreement
with the experimental values ($^{75}\nu_Q$)$_{\textrm{exp}}=$14.6
MHz and ($^{87}\nu_Q$)$_{\textrm{exp}}=$ 6.2 MHz and the
discrepancy represents an estimate of the accuracy of the DFT
calculation which is known to not properly account for the
electronic correlations.

%%%%%%%%%%%%%%%%%%%%%%%%%%%%%%
\begin{figure}[htpb]
\vspace{5cm} \includegraphics{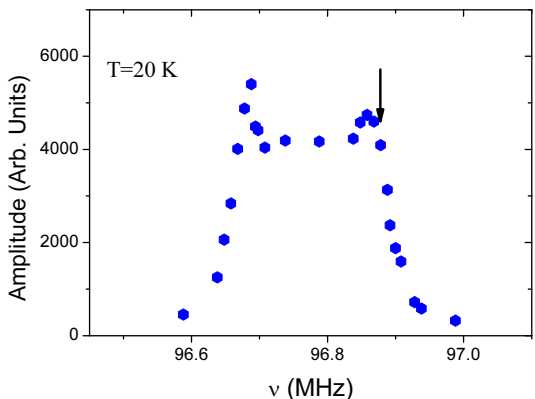} \caption{$^{87}$Rb NMR powder spectrum
in RbFe$_2$As$_2$ for the central $m_I= 1/2\rightarrow -1/2$
transition is shown for $T=20$ K. The arrow marks the position
where $^{87}$Rb NMR $1/T_1$ was measured.} \label{RbNMR}
\end{figure}
%%%%%%%%%%%%%%%%%%%%%%%%%%%%%%%%%

\section{III. NQR and NMR Spectra}

$^{75}$As NQR and $^{87}$Rb NMR spectra were derived by recording
the integral of the echo signal after a $\pi/2-\tau_e-\pi$ pulse
sequence as a function of the irradiation frequency. At a few
temperatures the $^{75}$As NQR spectra was also obtained by
merging the Fourier transforms of half of the echo recorded at
different frequencies but no relevant additional features appeared
in the spectra. We point out that any tiny amount of spurious
phases as FeAs and Fe$_2$As (not detected in X-ray diffraction)
will not affect the $^{75}$As NQR spectra since these materials
are magnetically ordered and the internal field shifts the
resonance frequency to much higher values. Also in FeAs$_2$ the
$^{75}$As NQR line is in a completely different frequency
range.\cite{FeAs2}

The narrow $^{87,85}$Rb NQR spectra were obtained from the Fourier
transform of half of the echo signal obtained after the same echo
pulse sequence. $^{87}$Rb NMR powder spectrum for the central
$m_I=1/2\rightarrow -1/2$ transition is displayed in
Fig.\ref{RbNMR}. The spectrum is fully compatible with the
$^{87}$Rb quadrupole frequency determined from the NQR spectra.

%%%%%%%%%%%%%%%%%%%%%%%%%%%%%%
\begin{figure}[h!]
\vspace{5cm} \includegraphics{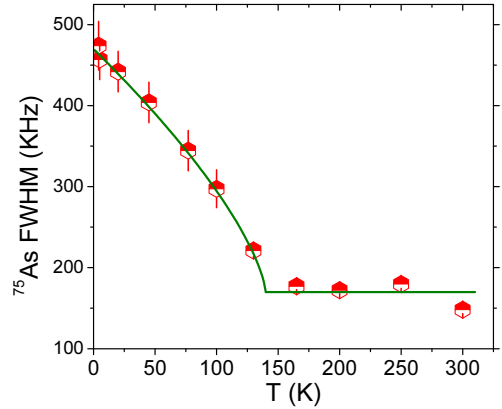} \caption{The temperature dependence of
$^{75}$As NQR spectrum full width at half maximum (FWHM) in
RbFe$_2$As$_2$ is reported.  The green solid line tracking the
order parameter is a phenomenological fit of $^{75}$As NQR FWHM
($\Delta\nu_Q$) with $\Delta\nu_Q= 300(1-(T/T_0))^\beta + 170$
KHz, with $T_0=140$ K and $\beta\simeq 0.7$.} \label{FWHMAs}
\end{figure}
%%%%%%%%%%%%%%%%%%%%%%%%%%%%%%%%%

\section{IV. Spin-spin relaxation rate $1/T_2$}

\begin{figure}[htpb]
\vspace{5cm} \includegraphics{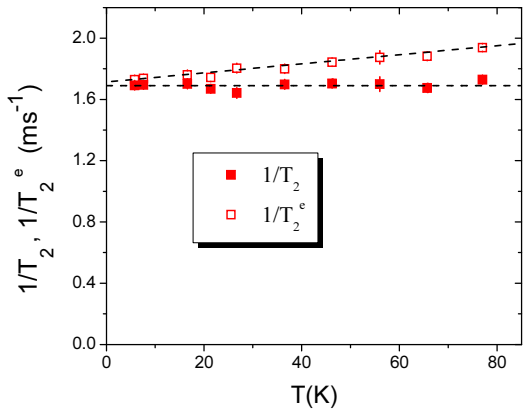} \caption{The temperature dependence of
$^{75}$As NQR $1/T_2^e$ (open squares) and $1/T_2$ (closed
squares), after Redfield correction, for RbFe$_2$As$_2$ is
reported. The dashed lines are guide to the eye.} \label{FigT2}
\end{figure}

$^{75}$As spin-spin relaxation rate $1/T_2$ was derived in NQR by
recording the decay of the echo amplitude $E(2\tau_e)$ after a
$\pi/2-\tau_e-\pi$ pulse sequence. The decay could be fit in
general with $E(2\tau_e)=E(0)exp(-2\tau_e/T_2^e)^\beta$ with
$\beta\simeq 1.6$. A value of $\beta$ lower than 2 and the slight
temperature dependence of $1/T_2^e$ (Fig.\ref{FigT2}) should be
associated with Redfield contribution to the relaxation
$1/T_{2R}$. Then one can write
$E(2\tau_e)=E(0)exp(-2\tau_e/T_2)^{\beta_2}exp(-2\tau_e/T_{2R})$,
with $1/T_2$ the spin-spin relaxation rate. In case of an
anysotropic spin-lattice relaxation rate, Walstedt \textit{et al}.
\cite{WC1995} calculated a general expression for ${1}/{T_{2R}}$.
In case of a nuclear spin $I=3/2$, with the $Z$ axes of the EFG
along the $c$ axes one should have:
\begin{equation}
\frac{1}{T_{2R}}=\frac{3}{T_{1}^{\parallel
c}}+\frac{1}{T_{1}^{\perp c}}
\end{equation}
where the symbols $\parallel$ and $\perp$ refers to the external
field orientation with respect to the crystallographic $c$ axis.
In particular, $1/T_{1}^{\parallel c}$ corresponds to $^{75}$As
NQR $1/T_1$. $1/T_{1}^{\perp c}$ was determined by assuming an
anisotropy in $1/T_1$ equal to the one found in electron-doped
BaFe$_2$As$_2$ \cite{Bossoni}. Once the data have been corrected
by Redfield contribution one finds that $^{75}$As $1/T_2$ is
temperature independent (Fig.\ref{FigT2}), with $\beta_2\simeq
1.8$. The deviation of $\beta_2$ from 2, as it is expected in the
case of nuclear dipolar interaction in a dense system
\cite{Abragam} is likely a consequence of the partial irradiation
of the NQR spectrum.
%%%%%%%%%%%%%%%%%%%%%%%%%%%%%%
%%%%%%%%%%%%%%%%%%%%%%%%%%%%%%%%%

Taking into account of $T_2$ corrections we have measured the
$T$-dependence of $^{75}$As NQR spectrum amplitude below 50 K and
did not observe any significant change.

\section{V. Nuclear spin-lattice relaxation rate $1/T_1$}

\begin{figure}[htpb]
\vspace{5cm} \includegraphics{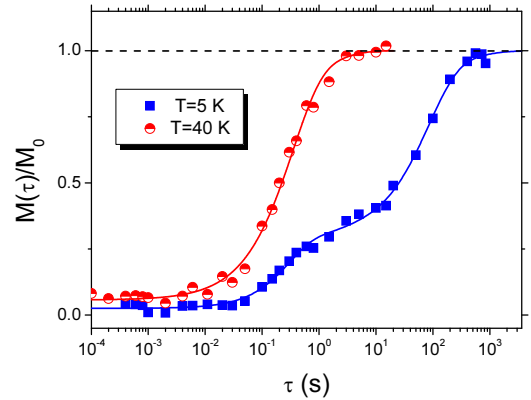} \caption{The recovery of $^{87}$Rb NMR
magnetization after a saturation pulse sequence, when the high
frequency peak of the NMR spectrum for the central $1/2\rightarrow
-1/2$ transition is irradiated, at $T= 40$ K (red circles) and at
$T= 5$ K (blue squares). The solid lines are the best fits
according to Eq. 6, for $T= 40$ K, and according to the same
recovery law, but considering two components, for $T= 5$ K. }
\label{T1RbNMR}
\end{figure}

$^{87}$Rb NMR $1/T_1$ was measured in a $H= 7$ Tesla magnetic
field by irradiating just the high frequency shoulder of the
powder spectrum of the central line shown in Fig.\ref{RbNMR},
corresponding to grains with the $c$-axes perpendicular to $\vec
H$. The recovery of $^{87}$Rb NMR central line magnetization after
a saturation recovery pulse sequence was fit according to
\begin{equation} \label{RbNMRrecovery}
M(\tau)=M_0\biggl( 1 - f(0.9 e^{-6\tau/T_1}+ 0.1
e^{-\tau/T_1)}\biggr) \,\,\, .
\end{equation}
The recovery is shown in Fig.\ref{T1RbNMR} and one observes,
similarly to what one finds in $^{75}$As NQR, two components
appearing at low temperature.

%%%%%%%%%%%%%%%%%%%%%%%%%%%%%%
%%%%%%%%%%%%%%%%%%%%%%%%%%%%%%%%%

The long component were measured just in NMR since in NQR the very
long $^{87}$Rb relaxations and the much lower signal intensity
make the measurements quite demanding. The fast component, the
only one present at $T>25$ K, was measured in NQR irradiating
either $^{87}$Rb $\pm 3/2\rightarrow \pm 1/2$ transition or
$^{85}$Rb $\pm 5/2\rightarrow \pm 3/2$. The recovery of nuclear
magnetization was fit according to the recovery laws expected for
a magnetic relaxation mechanism \cite{Mac}
\begin{equation} \label{87RbNQRrecovery}
M(\tau)=M_0\biggl( 1 - fe^{-3\tau/T_1}\biggr) \,\,\, ,
\end{equation}
for $^{87}$Rb and

%%%%%%%%%%%%%%%%%%%%%%%%%%%%%%
\begin{figure}[htpb]
\vspace{5.5cm} \includegraphics{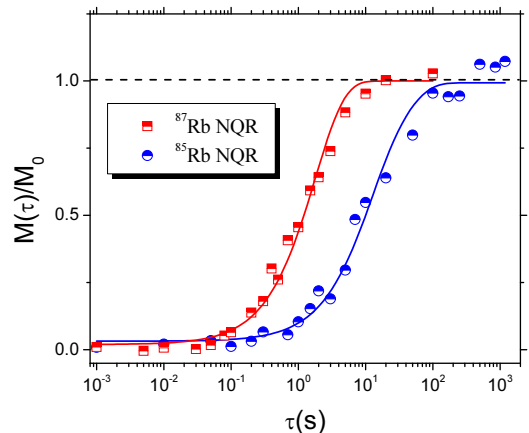} \caption{The recovery of $^{87}$Rb (red
squares) and of $^{85}$Rb (blue circles, $\pm 5/2\rightarrow \pm
3/2$ transition) magnetization in NQR after a saturation pulse
sequence, for $T= 4.2$ K. The solid lines are the best fits
according to Eqs. 7 and 8, respectively. Although the recovery law
(Eq.\ref{85RbNQRrecovery}) for $^{85}$Rb would imply a faster
recovery, it is clear that $^{87}$Rb relaxation is much faster as
it is expected for a magnetic relaxation mechanism. }
\label{T1RbNQR}
\end{figure}
%%%%%%%%%%%%%%%%%%%%%%%%%%%%%%%%%

\begin{equation} \label{85RbNQRrecovery}
M(\tau)=M_0\biggl( 1 - f(0.427 e^{-3\tau/T_1}+ 0.573 e^{-10
\tau/T_1})\biggr) \,\,\, ,
\end{equation}
for $^{85}$Rb. The ratio between the $1/T_1$ of the two nuclei for
the fast relaxing component was measured at a few selected
temperatures between 4 and 25 K and $^{87}(1/T_1)/^{85}(1/T_1)= 12
\pm 1$ (Fig.\ref{T1RbNQR}), in good agreement with the ratio
between the square of the gyromagnetic ratios of the two nuclei
$(^{87}\gamma/^{85}\gamma)^2= 11.485$, confirming the adequacy of
the recovery laws we have used to estimate $1/T_1$ and the fact
that the relaxation is driven by electron spin fluctuations.
Notice that if the relaxation was driven by EFG fluctuations
$1/T_1$ should scale with the square of the nuclear electric
quadrupole moment and one should have $^{87}(1/T_1)/^{85}(1/T_1)=
0.226$ a value about 50 times smaller than the experimental one.

%%%%%%%%%%%%%%%%%%%%%%%%%%%%%%
\begin{figure}[h!]
\vspace{5.7cm} \includegraphics{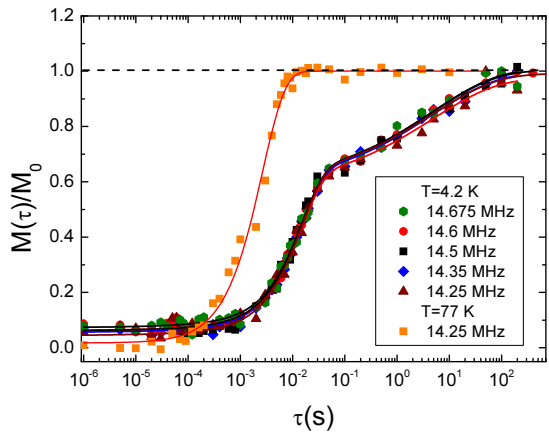} \caption{The recovery of $^{75}$As NQR
magnetization after a saturation pulse sequence for different
irradiation frequencies across the NQR spectrum, at $T= 4.2$ K.
The single exponential recovery law for $T= 77$ K is reported up
to times longer than $10^3 T_1$. The solid lines are best fits
according to Eq.2 in the main article.} \label{T1vsfreq}
\end{figure}
%%%%%%%%%%%%%%%%%%%%%%%%%%%%%%%%%

We have measured the frequency dependence of $^{75}$As $1/T_1$
across the NQR spectrum by decreasing the intensity of the
radiofrequency field so that we irradiated a width of about 30 KHz
in the spectrum. We found that the recovery laws did not change
significantly across the spectrum (Fig.\ref{T1vsfreq}). Namely,
$A_f$, $A_s$, $1/T_1^f$ and $1/T_1^s$ show a negligible frequency
dependence. It should be mentioned that $A_s$ and $A_f$ appear to
slightly depend on the thermal history (i.e. on the cooling rate),
an aspect that will be the subject of future studies. Finally, we
have checked that $A_s$, the amplitude of the slow relaxing
component, is zero above 20 K by recording the recovery up to more
than $10^{3}$ $T_1$ (Fig.\ref{T1vsfreq}).

\end{document}